\documentclass{iau}
\usepackage{graphicx}

\title[B-stars in the open cluster NGC~884] 
{An observational asteroseismic study of the pulsating B-stars in the open cluster NGC~884}

\author[S.~Saesen et al.]   
{S.~Saesen$^{1,2}$,
M.~Briquet$^{2,3,\thanks{Postdoctoral Researcher of the Fund for Scientific Research, Fonds de la Recherche Scientifique -- FNRS, Belgium}}$,
C.~Aerts$^{2,4}$,
A.~Miglio$^5$,
\and F.~Carrier$^2$}

\affiliation{$^1$Observatoire de Gen\`eve, Universit\'e de Gen\`eve, Switzerland
\\ email: {\tt sophie.saesen@unige.ch} \\[\affilskip]
$^2$Instituut voor Sterrenkunde, Katholieke Universiteit Leuven, Belgium\\
$^3$Institut d'Astrophysique et de G\'eophysique de l'Universit\'e de Li\`ege, Belgium\\
$^4$Department of Astrophysics, Radboud University Nijmegen, The Netherlands\\
$^5$School of Physics and Astronomy, University of Birmingham, UK}

\pubyear{2013}
\volume{301}  
\pagerange{xx--xx}
\jname{Precision Asteroseismology}
\editors{W.~Chaplin, J.~Guzik, G.~Handler \& A.~Pigulski, eds.}
\begin{document}

\maketitle

\begin{abstract}
Recent progress in the seismic interpretation of field $\beta$ Cep stars has resulted in improvements of the physical description in the stellar structure and evolution model computations of massive stars. Further asteroseismic constraints can be obtained from studying ensembles of stars in a young open cluster, which all have similar age, distance and chemical composition.  We present an observational asteroseismic study based on the discovery of numerous multi-periodic and mono-periodic B-stars in the open cluster NGC 884. Our study illustrates the current status of ensemble asteroseismology of a young open cluster.
\end{abstract}

\firstsection 
\section{Data}

We exploited the differential time-resolved multi-colour photometry of a selected field of NGC~884 presented in \cite[Saesen et al. (2010)]{Saesen10}. The photometry was gathered at 12~sites using 15~different instruments and resulted in almost 77500~CCD images and 92~hours of photo-electric data. The data span three observation seasons (2005 -- 2007) and the resulting precision of the light curves of the brighter stars is 5.7~mmag in $V$, 6.9~mmag in $B$, 5.0~mmag in $I$ and 5.3~mmag in $U$. We identified 75~periodic B-stars through an automated frequency analysis in the $V$ filter and focus on these stars in the current study. We derived effective temperature and luminosity values from absolute photometry in the seven Geneva filters and we combined them with literature values to get the position of the B-stars in the HR-diagram.

\section{Derivation of the pulsational properties of the B-type stars}

We performed a detailed frequency analysis with {\sc period04 }\cite[(Lenz \& Breger 2005)]{Lenz05}. We used a weighted and non-weighted frequency analysis for the $V$ light curve, and only a weighted analysis for $B$, $I$ and $U$. We considered a peak significant if it has a signal-to- noise ratio above 4. A classical prewhitening scheme was adopted to determine all significant frequencies present in a certain star. As a result, we keep 65~periodic B-stars of which 36~are mono-periodic, 16~are bi-periodic, 10~are tri-periodic, 2~are quadru-periodic and one star has 9~independent frequencies. The results of the multi-frequency fits in the different colours are presented in Table 4 of \cite[Saesen et al. (2013)]{Saesen13}.

Since none of the detected frequencies show any phase differences in the different filters, we restricted for the mode identification to the photometric amplitude ratios to determine the degree $\ell$, following the method of Dupret et al. (2003). For each pulsator, we selected stellar equilibrium models that fitted the observed position in the HR-diagram within a 2$\sigma$-error box. We then selected for each model and each degree $\ell$ between 0 and 4 the theoretical frequencies that fit the observed ones, taking into account rotational splitting. We confronted the theoretical amplitude ratios with those observed and made a mode identification by eliminating the degrees for which the theoretical amplitude ratios do not match the observed ones. We securely identified the degree for 12 of the 114 detected frequencies.

\section{Relations between pulsation and basic stellar parameters}

For the following analyses, we exclude three stars: one ellipsoidal binary and two non-member $\delta$~Sct stars. All other 62~B-stars can be considered members of NGC~884 based on photometric diagrams and membership studies in the literature. We did not find a correlation between the projected rotational velocity and the amplitude or frequency of the modes. The amplitudes of the oscillations, however, decrease as the frequency values increase. A comparison between the observed and theoretical frequency-radius relation allowed us to identify eight $\beta$~Cep stars in our sample, namely Oo~2246, Oo~2299, Oo~2444, Oo~2488, Oo~2520, Oo~2572, Oo~2601 and Oo~2694.

\section{Age estimate of the cluster}

The seismic properties of the selected $\beta$~Cep stars were confronted with those predicted for a dense full grid of stellar models \cite[(Briquet et al. 2011)]{Briquet2011}. We searched for a consistent cluster solution by requesting common cluster parameters, i.e., equal age and initial chemical composition, without specifying their values. We eliminated models based on the observed position of the stars in the HR-diagram and the observed frequency spectrum, allowing for rotational splitting in the first-order approximation. Imposing the identified degrees and measured frequencies of the radial, dipole and quadrupole modes of five $\beta$ Cep stars led to a seismic cluster age estimate of $\log(\mathrm{age/yr})=7.12-7.28$. This is fully compatible with the age estimate obtained from modeling of an eclipsing binary in the cluster ($\log(\mathrm{age/yr}) > 7.10$, \cite[Southworth et al. 2004]{Southworth04}) and illustrates the valuable alternative that cluster asteroseismology can offer for age determinations.

\section{Further details}
For more information, we refer the interested reader to \cite[Saesen et al. (2010, 2013)]{}.

\end{document}